\documentclass[a4paper,11pt]{article}
\usepackage[utf8x]{inputenc}
\usepackage[T1]{fontenc}
\usepackage{multicol}
\usepackage{float}
\usepackage{vmargin}
\usepackage{amsmath}
\usepackage{amsfonts}
\usepackage{amssymb}
\usepackage{amsthm}
\usepackage{bbold}
\usepackage{latexsym}
\usepackage{feynmf}
\usepackage{color}
\usepackage{tensor}
\usepackage{graphics}
\usepackage{graphicx}

\usepackage{vmargin}
\setpapersize{A4}
  \setmargins{2.5cm}{1cm}    %linker & oberer Rand
	     {16cm}{25cm}    %Textbreite und -hoehe
	     {12pt}{25pt}    %Kopfzeilenhoehe und -abstand
	     {0pt}{30pt}     %\footheight (egal) und Fusszeilenabstand

\DeclareMathOperator{\pa}{\partial}

\newcommand{\reell}{\mathbb{R}}

\begin{document}

\begin{center}
{\LARGE Conformal transformations of the S-matrix; $\beta$-function
identifies change of spacetime}
\vspace{0.5cm}

{\large Steffen Pottel, Klaus Sibold\\
Institut f\"ur Theoretische Physik\\
Universit\"at Leipzig\\
Postfach 100920\\
D-04009 Leipzig
Germany}
\end{center}

\begin{abstract}
First conformal transformations of the $S$-matrix  are derived in massless 
$\phi^4$-theory. Then it is shown that the anomalous transformations
can be rewritten as a symmetry once one has introduced a local coupling
and interprets the charge of the symmetry accordingly. By introducing
a suitable effective coupling on which the $S$-matrix depends one is able
to identify via the $\beta$-function an underlying new spacetime with
non-trivial conformal (flat) metric.
\end{abstract}

\section{Introduction}
It is well-known that conformal transformations realized in quantum field 
theories in four-dimensional spacetime as a rule are beset with anomalies.
Formulated in terms of Green functions (which are off-shell quantities)
they are parametrized by $\beta$-functions, associated with the anomalous
behaviour of interaction vertices and by $\gamma$-functions associated with 
anomalous dimensions of the fields. Exceptions were discovered mainly in 
the context of supersymmetric quantum field theories, notably  the 
famous $N=4$ super-Yang-Mills theory. The question arises which effects 
of these anomalies survive in physical quantities like $S$-matrix or Green
functions of physical operators. A first answer has been given quite some
time ago by Zimmermann \cite{Zimmermannphi4} for the dilatations in massless
$\phi^4$. In an axiomatic setting he has shown that the $S$-operator
scales with the $\beta$-function
\begin{equation} \kappa^2\partial_{\kappa^2} S =
                  \beta_\lambda\partial_\lambda S.\label{RGequat}
\end{equation}
This can be understood as the renormalization group equation for the 
$S$-operator. There is no contribution by $\gamma$. This result depends
crucially on the fact that the propagator can be shown to have a pole
at vanishing momentum with finite residue and therefore a field operator
exists whose propagator has a pole at zero momentum with residue one.  
Hence the appearence of $\gamma$ is a consequence of normalizing the field 
operator unphysically.\\
A first attempt to extend this result to the special conformal transformations for a massive $\phi^4$ theory has been undertaken in \cite{YongZhang}.
But the massless limit was not considered there, since the calculation
suffered from divergences, which were too difficult to control.
In the present paper we work directly in the massless limit by using
the Bogoliubov-Para\v siuk-Hepp-Zimmermann-Lowenstein (BPHZL) subtraction 
scheme. Assuming that the same normalization conditions hold as the ones 
Zimmermann used we find that the special conformal transformations change
the $S$-operator according to 
\begin{equation}\label{confSop}
i \left[K_\mu,S\right] = \lim \int dx\,
 2x_\mu\beta_\lambda\frac{\delta}{\delta\lambda(x)} S
\end{equation}
``$\lim$'' refers to {\sl constant} coupling $\lambda$, because 
in the course of deriving this result the coupling $\lambda$ had been 
generalized to be a function of 
spacetime, i.e.\ to vary with $x$. Here, too no effect of $\gamma$ shows up.
By putting in (\ref{confSop}) the right-hand-side to the left one can
interpret the new equation as the expression for a symmetry, where
the notion of charge is extended to include the external field $\lambda$.
Yet another possibility of using (\ref{confSop}) will lead to the
identification of an underlying spacetime which is still flat, but 
has a non-trivial conformal metric. Hence one has an interacting theory
which lives on a spacetime which is not
Minkowski and satisfies the axioms in the sense of perturbation theory.

The paper is organized as follows. In Sec.\ II we recapitulate a few facts
on conformal transformations for constant and for $x$-dependent coupling
$\lambda$. In particular we present the Ward identity (WI) for
dilatations and special
conformal transformations of the Green functions after the anomalies have 
been absorbed. In Sec.\ III we derive the transformation laws for the 
$S$-operator. In Sec.\ IV these transformation laws will be reinterpreted
on a transformed space (by dilatations and special conformal transformations
respectively). Sec.\ V contains discussion and conclusions.

\section{Dilatations and special conformal transformations of the Green functions}
\subsection{The classical approximation}
The classical action
\begin{equation}
\Gamma_{\hbox{\rm cl}}=\int d^4x (\frac{1}{2}\partial\phi\partial\phi
-\frac{1}{4!}\lambda \phi^4)
\end{equation}
is invariant under dilatations and special conformal transformations
\begin{align}\label{classvar}
\delta^D(x,d=1)\phi &= (1+x^\mu\partial_\mu)\phi \\
\delta^K_\mu(x,d=1) \phi &= (2x_\mu x^\lambda-
      \eta\indices{_\mu^{\lambda}} x^2)\partial_\lambda \phi +2 x_\mu \phi
\end{align}
Interpreting the classical action as the tree graph approximation of
the generating functional $\Gamma$ for vertex functions (one-particle-irreducible
Green functions) 
\begin{equation}
\Gamma_{\hbox{\rm cl}} = \Gamma^{(0)}
\end{equation}
(loop number = zero),
one can express this invariance as a WI
\begin{align}\label{Wclass}
W^D\Gamma^{(0)} &= 0,\\
W^K_\mu\Gamma^{(0)} &= 0,
\end{align}
where 
\begin{align} \label{WIopcl}
W^D &\equiv -i\int d^4x  \delta^D\phi \frac{\delta}{\delta\phi},\\
W^K_\mu &\equiv -i\int d^4x  \delta^K\phi \frac{\delta}{\delta\phi}.
\end{align}

In higher orders these WI's will be broken by anomalies. In order
to deal with those it is convenient to promote the coupling constant
$\lambda$ to an external field $\lambda(x)$, for then one can generate
non-integrated vertices by differentiation with respect to $\lambda(x)$,
a process which in the BPHZL renormalization scheme is under control
by the action principle. It is also very useful to construct respective
currents as $x$-moments of the improved energy-momentum tensor. This
can be realized by starting from suitable WI operators (``contact terms'').
 \begin{align}
 \tilde w^T_\nu & \equiv \pa_\nu \phi - \frac{1}{4} \pa_\nu \left( \phi
                    \frac    {\delta}{\delta \phi} \right) +
		    \pa_\nu \lambda \frac{\delta}{\delta \lambda} \\
  W^D[\phi,\lambda] & \equiv -i\int dx\, x^\mu \, \tilde w^T_\mu \\
  W^K_\mu [\phi,\lambda] & \equiv -i\int (2x_\mu x^\nu -
                     \eta_\mu^{\,\,\,\nu} x^2) \tilde w^T_\nu 
 \end{align}

Here the external field $\lambda(x)$ has been assigned vanishing
canonical dimension and assumed to be $\lambda(x) \in
\mathcal{S}(\mathbb{R}^4)$ .

\subsection{Higher orders}
For quantization and calculation of higher orders we employ the
BPHZL renormalization scheme which first of all means introducing
an auxiliary mass term 
\begin{equation}\label{massterm}
\Gamma^{\textnormal{mass}} = \left[\int \! d^4x\, \Big(\! -\frac{1}{2}M^2(s-1)^2 
\phi^2\Big)\right]^4_4,
\end{equation}
which results into a free propagator $\Delta_c$  
\begin{equation} \Delta_c(x)=\frac{i}{(2\pi)^4}\int d^4p 
                  \frac{e^{-ipx}}{p^2-M^2(s-1)^2+i\varepsilon_{\textnormal Z}}.\label{propagator} 
\end{equation}		  
 \[\varepsilon_{\textnormal{Z}} = \varepsilon (M^2(s-1)^2 + {\bf p}^2)\]
is Zimmermanns epsilon which yields Euclidian minorants and majorants for
the momentum space integrals of Feynman diagrams.
The variables $s$ and $s-1$ participate in the subtractions like the
external momenta and Zimmermanns epsilon leads to absolute convergence
once subtractions have been properly performed. Non-trivial quantum
corrections can show up when one wants to go
to the massless limit $s=1$ which is possible only in expressions where
$s-1$ appears outside of normal products and thus does no longer participate
in the subtractions. The relation between such normal
products differing only in the position of the $s-1$ factors is given by
an identity (due to Zimmermann) to which all non-naive deviations from, say 
symmetry in the quantum theory can be traced back.\\
The next ingredient is $\Gamma_{\textnormal{eff}}$
from which Feynman diagrams follow. In the BPHZL renormalization scheme,
where $\Gamma_\mathrm{eff} = \Gamma_\mathrm{free}
 + \int \mathcal{L}_\mathrm{int}$, $\Gamma_\mathrm{eff}$ is to
 be understood as a normal product with infrared and
ultraviolet subtraction degree four.
  In the case of local coupling $\lambda(x)$,
external field $h_{\mu\nu}(x)$ (to which the energy-momentum tensor couples) and
quantum field $\phi$ it is given by \cite{KrausSibold1}, 
\cite{KrausSibold2} 
\begin{eqnarray}\label{gammaeff}
\Gamma_{\textnormal{eff}}(\phi,\lambda,h)&=& \sum_{n=0}^{\infty}\int \left( -\frac{1}{2}z^{(n)}I_l^{(n)}
                   -\frac{1}{2}M^2(s-1)^2I_M\delta_{n,0} \right. \nonumber \\
		  &\phantom{=}& \left. -\frac{1}{4!}\rho^{(n)}I_4^{(n+1)}
         +\frac{1}{2}\hat{c}^{(n)}I_c^{(n)}+\tilde{z}^{(n)}I_1^{(n)}
         +z_{\lambda}^{(n)}I_{\lambda}^{(n)} \right)
\end{eqnarray}
with the basis of $(4,4)$- insertions 
\begin{eqnarray}\label{basis}
I_M &=& (-g)^{1/4}\phi^\\[5pt]
I_l^{(n)}&=& (-g)^{3/8}g^{\mu\nu}\lambda^n\phi
   (\partial_{\mu}\partial_{\nu}-\Gamma_{\mu\nu}^{\nu'}\partial_{\nu'})
                                ((-g^{-1/8}\phi)\\[5pt]
I_4^{(n)}&=& \lambda^n\phi^4\\[5pt]
I_c^{(n)}&=& (-g)^{1/4}\lambda^n R\phi^2\\[5pt]
I_1^{(n)}&=&(-g)^{1/2}g^{\mu\nu}\lambda^{n-1}
                          \partial_{\mu}\lambda\partial_{\nu}((-g)^{-1/4}\phi^2)\\[5pt]
I_{\lambda}^{(n)}&=&(-g)^{1/4}g^{\mu\nu}\lambda^{n-2}
                          \partial_{\mu}\lambda\partial_{\nu}\lambda\phi^2\\[2pt]
\hat{I}^{(n)}_k&=&(-g)^{1/2}g^{\mu\nu}\frac{1}{2}(\partial_{\mu}\partial_{\nu}-
          \Gamma_{\mu\nu}^{\nu'}\partial_{\nu'})((-g)^{-1/4}\lambda^n\phi^2)\\
\hat{I}^{(n)}_2&=& (-g)^{1/2}g^{\mu\nu}\frac{1}{n}(\delta_{\mu}^{\nu'}\partial_{\nu}
-\Gamma_{\mu\nu}^{\nu'})((-g)^{-1/4}\phi^2\partial_{\nu'}\lambda^n)
\end{eqnarray}
Here $g=\det(g_{\mu\nu})$, $R$ is the curvature scalar, and $\Gamma^{\rho}_{\mu\nu}$
the Christoffel symbol in the usual conventions.
In the following this curved background spacetime will not be needed
explicitly -- all curvature dependent terms will vanish and the Christoffel
symbols are constant in the case of dilatations and special conformal
transformations. 
For the derivation of the subsequent WI's they are however crucial
and similarly for considerations of other observables than the $S$-matrix
studied in the sequel.\\
As far as normalization conditions are concerned we spell out explicitly only
those relevant for $h_{\mu\nu}=0$ and $\lambda={\hbox{\rm const}}$
\[\partial_{p^2}\Gamma_{\phi\phi}|_{\stackrel{p^2=-\kappa^2}{ s=1}}=1,  \quad
\Gamma_{\phi\phi\phi\phi}|_{\stackrel{p=p(symm)}{s=1}} = -\lambda.\]  
The subtraction scheme implies that $\Gamma_{\phi\phi}$ vanishes at $p=0, s=1$.
(For the complete list we refer to \cite{KrausSibold2}.)

Going through some algebraic analysis which is based on the presence of
$h_{\mu\nu}$ and using consistency conditions one finds
(s.\cite{KrausSibold2}) that in all higher orders of perturbation theory 
the following broken WI's hold at $s=1$

\begin{align}
W^D\Gamma[\phi,\lambda]=&-i\sum^\infty_{k=1}\limits \left( \int dx\; \hat 
\beta^{(k)}_\lambda
 \lambda^{k+1}(x) \frac{\delta}{\delta \lambda(x)} - \int dx\, \hat \gamma^{(k)} 
\lambda^k(x) \phi \frac{\delta}{\delta \phi}  \right)\Gamma[\phi,\lambda] \\
 W^K_\mu\Gamma[\phi,\lambda]=&-i\sum^\infty_{k=1}\limits \left( \int dx\; 2x_\mu\,
 \hat \beta^{(k)}_\lambda \lambda^{k+1}(x) \frac{\delta}{\delta \lambda(x)} - \int 
dx\, 2x_\mu \hat \gamma^{(k)} \lambda^k(x) \phi \frac{\delta}{\delta \phi} \right) 
\Gamma[\phi,\lambda]\\
 \beta_\lambda (x) & \equiv \sum^\infty_{k=1}\limits \hat \beta_\lambda^{(k)} 
                                                     \lambda^{k+1} (x) \\
 \gamma (x) & \equiv \sum^\infty_{k=1} \hat \gamma^{(k)} \lambda^k (x)
\end{align}

It is then tempting to absorb the right-hand-sides of the WI's into new 
WI-operators and to generate homogeneous WI's
\begin{align} 
\hat W^D [\phi,\lambda] & \equiv W^D [\phi,\lambda] + i \sum^\infty_{k=1}\limits \left( \int dx\; \hat \beta^{(k)}_\lambda \lambda^{k+1}(x) \frac{\delta}{\delta \lambda(x)} - \int dx\, \hat \gamma^{(k)} \lambda^k(x) \phi \frac{\delta}{\delta \phi}  \right) \\
\hat W^D \Gamma[\phi,\lambda] & =0 \label{homDWI}\\
 \hat W^K_\mu [\phi,\lambda] & \equiv W^K_\mu[\phi,\lambda] + i\sum^\infty_{k=1}\limits 
\left( \int dx\; 2x_\mu\, \hat \beta^{(k)}_\lambda \lambda^{k+1}(x) \frac{\delta}{\delta \lambda(x)} - \int dx\, 2x_\mu \hat \gamma^{(k)} \lambda^k(x) \phi \frac{\delta}{\delta \phi} \right) \\
\hat W^K_\mu \Gamma[\phi,\lambda] & =0
\end{align}
A good reason to proceed into this direction stems from the remarkable fact, 
already stated in \cite{KrausSibold2}, that these
new WI-operators fit to the moment construction and satisfy the conformal algebra.
The derivatives with respect to $\lambda (x)$ generate insertions hence one can 
in the limit of constant $\lambda$ understand the terms added
to the original WI operators as non-linear field transformations which complete the
linear transformations to formally true symmetries of the system. We may thus expect
conserved currents and associated charges operating on the Hilbert space
of the theory, involving however the external field $\lambda$.

To pave the way to the charge operators one needs the WI's formulated on 
the generating functional of the general Green functions.\\
In a first step one goes over to the connected Green functions by Legendre 
transformation
\begin{eqnarray}\label{Legtransf}
Z_c(J) &=& \Gamma(\phi) + \int J\phi\\
-J &=& \frac{\delta\Gamma}{\delta\phi},
\end{eqnarray}
and then in a second step to general Green functions
\[Z=e^{iZ_c}.\]
The WI's become 
\begin{align}
 \hat W^D [J,\lambda] \equiv&  -\int dx\, J(x) \delta^D (x,d=1) \frac{\delta}{\delta J(x)} + \int dx\, \delta^D(x,d=0) \lambda(x) \frac{\delta}{\delta \lambda (x)}\\
 & - \sum^\infty_{k=1}\limits \left( \int dx\, \hat \beta^{(k)}_\lambda \lambda^{k+1}(x) \frac{\delta}{\delta \lambda(x)} + \int dx\, \hat \gamma^{(k)} \lambda^k(x) J(x) \frac{\delta}{\delta J(x)} \right)\\
\hat W^D Z[J,\lambda] =& 0 \label{WIDoff}\\ 
 \hat W^K_\mu [J,\lambda] \equiv& -\int dx\, J(x) \delta^K_\mu (x,d=1) \frac{\delta}{\delta J(x)} + \int dx\, \delta^K_\mu(x,d=0) \lambda(x) \frac{\delta}{\delta \lambda (x)} \\
 & - \sum^\infty_{k=1}\limits \left( \int dx\; 2x_\mu\, \hat \beta^{(k)}_\lambda \lambda^{k+1}(x) \frac{\delta}{\delta \lambda(x)} + \int dx\, 2x_\mu \hat \gamma^{(k)} \lambda^k(x) J(x) \frac{\delta}{\delta J(x)} \right)\\
\hat W^K_\mu Z[J,\lambda] =& 0 \label{WIKoff}
\end{align}
As above for the one-particle-irreducible Green functions they express
the formal invariance of general Green functions under these generalized,
non-linear transformations. The WI's hold at $s=1$, the massless theory.
The respective Green functions exist as Lorentz invariant distributions
for non-exceptional momenta, the vertex functions as Lorentz invariant
functions for non-exceptional momenta. At values $s\not=1$ the WI's are
broken by soft mass contributions vanishing in the deep asymptotic
Euclidian region. As long as the coupling is local there is no infrared
problem anyway.

\section{Transformation laws for the $S$-operator}
Starting point for the subsequent analysis is the $S$-operator as defined by
\begin{align}
     S & = \,\colon\!\Sigma\colon Z_{|J=0} \\
\Sigma & = \exp \left\{ \int dx\, dy \, \phi_\mathrm{in}(x)
              \, r^{-1}\, K(x-y)\frac{\delta}{\delta J(y)} \right\} \\
  Z[J] & = \frac{\exp \{i \int \mathcal{L}_\mathrm{int}
                      (\frac{\delta}{i \delta J})\} \exp 
     \{\frac{1}{2}\int dx\, dy\, iJ(x) \Delta_c(x-y) iJ(y)\}}{[\exp \{i \int 
      \mathcal{L}_\mathrm{int}(\frac{\delta}{i \delta J})\} 
      \exp \{\frac{1}{2}\int dx\, dy\, iJ(x) \Delta_c(x-y) iJ(y)\}]_{|J=0}}.
\end{align}

Here, as above, $Z[J]$ denotes the generating functional for general
Green functions,
$\phi_\mathrm{in}(x)$ the asypmtotic field (which is free), $r$ the wave function 
renormalization constant and $K(x-y)= \Box + M^2(s-1)^2$ the inverse wave operator 
\cite{ItzyksonZuber}. In the massless $\lambda \phi^4$-theory we have

\begin{equation}
 \Sigma = \exp \left\{ \int dx\, \phi_\mathrm{in}(x)\, r^{-1}\, \Box_x 
                             \frac{\delta}{\delta J(x)} \right\}
\end{equation}

$\Sigma$ amputates the external legs and when evaluating the integral, $\phi_\mathrm{in}$ 
puts the external momenta on the mass shell.

Suppose now that a WI operator $W^A$ generates a symmetry $W^A Z(J)=0$, then 
one can
apply a standard method (s.\ \cite{BecchiLesHouches}, \cite{Kugobook}) to find
a symmetry of the $S$-operator in Hilbert space by establishing the following 
identity via integration by parts:
\begin{equation}\label{chargeScom}
 [W^A, :\!\Sigma:] Z_{|J=0} = [Q^A, \,:\!\Sigma:] Z_{|J=0}
\end{equation}

Here the left-hand-side is a calculation in the functional space, whereas
the right-hand-side is a calculation in Hilbert space with $Q^A$
representing the charge operator generating the transformation in question
\begin{align}
 i[Q^A,\phi_\mathrm{in}(x)] = \delta^A \phi_\mathrm{in}(x).
 \label{eq: becchi}
\end{align}
In simple cases the left-hand-side of (\ref{chargeScom}) vanishes because in the first 
contribution to
the commutator one uses the WI, in the second contribution one uses that at $J=0$
the WI operator vanishes. The right-hand-side represents the commutator
of the charge with the $S$-operator which is thus zero, hence the $S$-operator
is symmetric with respect to the transformation generated by $A$. In the
present case this holds e.g.\ true for the Lorentz transformations and the 
translations. For the dilatations and special conformal transformations
the situation is however more complicated: the $M^2(s-1)^2$-terms in $\Sigma$ disturb
the relation (\ref{chargeScom}), the reason for working with a $\Sigma$ at vanishing 
mass, i.e.\ at $s=1$ within $\Sigma$ and taking the propagators of the external legs also at $s = 1$.

\subsection{Dilatations}

Taking into account the remark at the end of the preceding paragraph we calculate the
commutator of $\Sigma$ at $s=1$ with $\hat W^D$ and put also in the external 
legs of the Green functions $s=1$. We obtain (Wick-dots omitted)

\[
 \hat W^D(\Sigma Z)_{|J=0,s=1}  = \left( \int dx\, \delta^D(x,d=0) \lambda(x) \frac{\delta}{\delta \lambda (x)} - \sum^\infty_{k=1}\limits \int dx\ \hat \beta^{(k)}_\lambda \lambda^{k+1}(x) \frac{\delta}{\delta \lambda(x)} \right) S \label{eq: WS},\]
where we performed as indicated the limit $J=0$ in the WI operator.
This is the non-trivial contribution of the commutator in the functional space
and represents eventually the breaking of the dilational symmetry.\\
The commutator
\begin{align}
 [\hat W^D, \Sigma] Z_{|J=0,s=1}  &= \Bigg\{ \int dx \, \phi_\mathrm{in}(x) \, r^{-1} \, \Box_x \, \delta^D (x,d=1)\frac{\delta}{\delta J(x)} \label{eq: delta}\\
  &\quad - \int dy\, \delta^D(y,d=0)\lambda(y) \int dx \, \phi_\mathrm{in}(x) \, r^{-1} \, \left( \frac{\delta}{\delta \lambda (y)} \ln(r_x) \right) \, \Box_x \, \frac{\delta}{\delta J(x)} \label{eq: lambda}\\
  &\quad + \sum^\infty_{k=1}\limits \int dy\, \hat \beta^{(k)}_\lambda \lambda^{k+1}(y) \int dx \, \phi_\mathrm{in}(x) \, r^{-1} \, \left( \frac{\delta}{\delta \lambda (y)} \ln(r_x) \right) \, \Box_x \,\frac{\delta}{\delta J(x)}\label{eq: beta}\\ 
 & \quad + \sum^\infty_{k=1}\limits \int dx\, \phi_\mathrm{in}(x) \, r^{-1} \, \Box_x \, \left(\hat \gamma^{(k)} \lambda^k(x) \frac{\delta}{\delta J(x)} \right) \Bigg\} \Sigma Z_{|J=0,s=1} \label{eq: gamma}
 \end{align}
is to stand for the commutator of the charge operator $D^\mathrm{op}$ with
the $S$-operator. Due to the special $s=1$ prescription (\ref{eq: delta}) becomes
\begin{align}
 \int dx \, \phi_\mathrm{in}(x) \, r^{-1} \, \Box_x \, \delta^D (x,d=1)\frac{\delta}{\delta J(x)} = - \int dx \, \delta^D (x,d=1)\phi_\mathrm{in}(x) \, r^{-1} \, \Box_x \, \frac{\delta}{\delta J(x)}.
\end{align}
If we now perform the limit $\lambda(x) \rightarrow \lambda$ at least the terms with derivatives of the coupling will vanish, especially $\delta^D(x,d=1)\lambda(x)$. Hence (\ref{eq: lambda}) will not contribute.
But similarly
\begin{align}
 \hat W^D [J,\lambda] (\Sigma Z)_{|J=0,s=1} = - \sum^\infty_{k=1}\limits \int dx\; \hat \beta^{(k)}_\lambda \lambda^{k+1}(x) \frac{\delta}{\delta \lambda(x)} S
\end{align}

In addition we use a statement to be found in \cite{Zimmermannphi4}.
Zimmermann proved there that in a
specific domain of the coupling constant, i.e.\ between two stationary points, 
it is possible to normalize on-shell even for the massless theory and the 
S-matrix elements exist. 
We choose the lowest possible regime (smallest coupling) in order to fulfil the requirements of 
perturbation theory and with the on-shell normalization conditions for the
propagator and the collinear point normalization for the coupling we are able to 
set the wave function renormalization constant (power series in $\lambda$) 
equal to one and find
\begin{align}
 \ln r_x = \ln 1 = 0
\end{align}
Hence (\ref{eq: lambda}) and (\ref{eq: beta}) vanish. (The transition to
the collinear point normalization for the coupling is only a finite
renormalization, compatible with the subtraction scheme, maintaining the
WI; the propagator
normalizations of \cite{Zimmermannphi4} and ours agree anyway.)

Zimmermann also stated that the anomalous dimension $\gamma =0$ after 
normalizing on-shell. We show this in an explicit calculation for every order.
Inserting

\begin{align}
 \phi_\mathrm{in}(x) & = \int \frac{d^3q}{(2\pi)^{\frac{3}{2}} \sqrt{2\omega_q}} (a_q \, e^{-iqx} + a^\dagger_q \, e^{iqx}) \\
 \lambda^k(x) & = \int \frac{d^4p}{(2\pi)^4} e^{-ipx} \Lambda(p) \\
 \Delta_c(x-y) & = \int \frac{d^4k}{(2\pi)^4} e^{-ik(x-y)} \frac{i}{k^2+i\epsilon_Z}
\end{align}

into (\ref{eq: gamma}), translating $\Box_x$ into Fourier space and performing the coordinate space integration we get

\begin{align}
 & \int dx\, \phi_{\mathrm{in}}(x) \, \Box_x \, \left( \lambda^k(x) \Delta_c(x-y)\right) \\
 & = \int  \frac{d^3q}{(2\pi)^{\frac{3}{2}} \sqrt{2\omega_q}} \frac{d^4p\, d^4k}{(2\pi)^4} e^{iky} (a_q\, \delta^{(4)}(q+k+p) + a^\dagger_q\, \delta^{(4)}(k+p-q) ) \frac{-(k+p)^2}{k^2+i\epsilon_Z}  \Lambda(p) \\
 & = \int  \frac{d^3q}{(2\pi)^{\frac{3}{2}} \sqrt{2\omega_q}} \frac{d^4p}{(2\pi)^4} \left( a_q \, e^{-i(p+q)y} \frac{-q^2}{(q+p)^2+i\epsilon_Z} + a^\dagger_q e^{i(p-q)y} \frac{-q^2}{(q-p)^2+i\epsilon_Z} \right)  \Lambda(p) \\
 & = 0
\end{align}

since $\phi_\mathrm{in}(x)$ is the asymtotic (free) field of the theory 
and sits on-mass-shell, i.e.\ $q^2 = 0$. In addition we know that the 
denominators lead to well-defined distributions and do not contain any ``hard''
singularities:

\begin{align}
 \lambda \in \mathcal{S}(\reell^4) & \Rightarrow \lambda^k \in \mathcal{S}(\reell^4) \Rightarrow \Lambda \in \mathcal{S}(\reell^4) 
\end{align}

 Collecting all arguments we obtain 

\begin{align}
 \int dx \, \delta^D (x,d=1)\phi_\mathrm{in}(x) \, r^{-1} \, \Box_x \, \frac{\delta}{\delta J(x)} \Sigma Z_{|J=0,s=1} = \sum^\infty_{k=1}\limits \int dx\;  \hat \beta^{(k)}_\lambda \lambda^{k+1}(x) \frac{\delta}{\delta \lambda(x)} S
\end{align}

 Recalling (\ref{eq: becchi}) and (\ref{eq: delta}) we have 

\begin{align}\label{dilcomS}
 i[D^{\textnormal{op}},\, S] = \lim_{\lambda\rightarrow\mathrm{ const}} \sum^\infty_{k=1}\limits \int dx\, \hat \beta^{(k)}_\lambda \lambda^{k+1}(x) \frac{\delta}{\delta \lambda(x)} S
\end{align}

where $D^{\textnormal{op}}$ is the same charge of the dilatations as in the case of 
constant coupling. \\
Due to
\begin{equation}\kappa^2\partial_{\kappa^2}\Gamma = iW^D \Gamma
\end{equation}
the commutator (\ref{dilcomS}) is nothing but the renormalization group equation 
(\ref{RGequat}) for the $S$-operator. Zimmermann derived this renormalization
group equation axiomatically under the assumption that perturbation theory 
and the nonperturbative theory are related to each other in the sense of an 
asymptotic series where all derivatives on, say Green functions etc., with 
respect to the coupling in the nonperturbative setting go over to the 
perturbative regime for small enough coupling. For the present perturbative 
derivation of the same relation this
implies that after one has arranged the on-shell normalization conditions
for the propagator (residue = 1, pole at p = 0) and collinear normalization for the coupling, maintaining the scheme independent WI's 
cancellations of potential infrared divergences occur in sums of diagrams
which might not be easily seen. In the one-loop approximation, where
only one diagram contributes there must not occur an infrared divergence.
This is indeed true as we show in the appendix.

\subsection{Special conformal transformations}
For the special conformal transformations we can follow the derivation of the
preceding subsection line by line.

\begin{align}
 [\hat W^K_\mu, \Sigma] Z_{|J=0,s=1} & = \hat W^K_\mu (\Sigma Z)_{|J=0,s=1} - \Sigma (\hat W^K_\mu Z)_{|J=0,s=1} \\
 \hat W^K_\mu (\Sigma Z)_{|J=0,s=1} & = \left( \int dx\, \delta^K_\mu(x,d=0) \lambda(x) \frac{\delta}{\delta \lambda (x)} - \sum^\infty_{k=1}\limits \int dx\; 2x_\mu\, \hat \beta^{(k)}_\lambda \lambda^{k+1}(x) \frac{\delta}{\delta \lambda(x)} \right) S \label{eq: WS}\\
 [\hat W^K_\mu, \Sigma] Z_{|J=0,s=1} & = \Bigg\{ \int dx \, \phi_\mathrm{in}(x) \, r^{-1} \, \Box_x \, \delta^K_\mu (x,d=1)\frac{\delta}{\delta J(x)} \label{eq: deltaK}\\
 & \quad - \int dy\, \delta^K_\mu(y,d=0)\lambda(y) \int dx \, \phi_\mathrm{in}(x) \, r^{-1} \, \left( \frac{\delta}{\delta \lambda (y)} \ln(r_x) \right) \, \Box_x \, \frac{\delta}{\delta J(x)} \label{eq: lambdaK}\\
 & \quad + \sum^\infty_{k=1}\limits \int dy\; 2x_\mu\, \hat \beta^{(k)}_\lambda \lambda^{k+1}(y) \int dx \, \phi_\mathrm{in}(x) \, r^{-1} \, \left( \frac{\delta}{\delta \lambda (y)} \ln(r_x) \right) \, \Box_x \,\frac{\delta}{\delta J(x)} \label{eq: betaK}\\
 & \quad + \sum^\infty_{k=1}\limits \int dx\, \phi_\mathrm{in}(x) \, r^{-1} \, \Box_x \, \left( 2x_\mu \hat \gamma^{(k)} \lambda^k(x) \frac{\delta}{\delta J(x)} \right) \Bigg\} \Sigma Z_{|J=0,s=1} \label{eq: gammaK}
 \end{align}

   Then (\ref{eq: deltaK}) becomes

\begin{align}
 \int dx \, \phi_\mathrm{in}(x) \, r^{-1} \, \Box_x \, \delta^K_\mu (x,d=1)\frac{\delta}{\delta J(x)} = - \int dx \, \delta^K_\mu (x,d=1)\phi_\mathrm{in}(x) \, r^{-1} \, \Box_x \, \frac{\delta}{\delta J(x)}
\end{align}

and we arrive also at

\begin{align}
 \hat W^K_\mu [J,\lambda] (\Sigma Z)_{|J=0,s=1} \rightarrow \hat W^K_\mu [0,\lambda] (\Sigma Z)_{|J=0,s=1} = \hat W^K_\mu [0,\lambda] S
\end{align}

In the limit of constant $\lambda$, $\lambda(x) \rightarrow \lambda$, at least 
the terms with derivatives of the coupling vanish, in particular
 $\delta^K_\mu(x,d=1)\lambda(x)$. Hence (\ref{eq: lambdaK}) will not contribute
and also

\begin{align}
 \hat W^K_\mu [J,\lambda] (\Sigma Z)_{|J=0,s=1} = - \sum^\infty_{k=1}\limits \int dx\; 2x_\mu\, \hat \beta^{(k)}_\lambda \lambda^{k+1}(x) \frac{\delta}{\delta \lambda(x)} S.
\end{align}

With the on-shell normalization conditions and in the limit of constant $\lambda$, 
as above
\begin{align}
 \ln r_x = \ln 1 = 0.
\end{align}

Finally we obtain

\begin{align}
 \int dx \, \delta^K_\mu (x,d=1)\phi_{\textnormal{in}}(x) \, r^{-1} \, \Box_x \, \frac{\delta}{\delta J(x)} \Sigma Z_{|J=0,s=1} = \sum^\infty_{k=1}\limits \int dx\; 2x_\mu\, \hat \beta^{(k)}_\lambda \lambda^{k+1}(x) \frac{\delta}{\delta \lambda(x)} S,
\end{align}
which is nothing but

\begin{align}\label{confcomS}
 i[K_\mu,\, S] = \sum^\infty_{k=1}\limits \int dx\; 2x_\mu\, \hat \beta^{(k)}_\lambda \lambda^{k+1}(x) \frac{\delta}{\delta \lambda(x)} S
\end{align}

where $K_\mu$ is the same charge of the special conformal transformations as 
in the case of constant coupling.

Here, in the conformal case we cannot dispose over a renormalization group 
argument if we wish to relate our perturbative result to a possible 
non-perturbative one, there is however an algebraic argument available. 
The WI operator $\hat W^K_\mu$
is via the moment construction related to $\hat  W^D$ and satisfies with it
the usual commutator relation
\[ \left[\hat W^D,\hat W^K_\mu\right] = i \hat W^K_\mu\]
to all orders of perturbation theory. Therefore we expect that this relation
is also valid in the nonperturbative theory and maintained in the sense of an
asymptotic expansion. Hence we expect here the same cancellations of
potential infrared
divergences to take place as in the case of dilatations.\\
In any case the off-shell relations (\ref{WIDoff}) and (\ref{WIKoff})
for the Green functions hold.

\subsection{Conserved transformations}
In this subsection we round up the discussion ``symmetry versus anomalies''
by giving an interpretation of
(\ref{dilcomS}) and (\ref{confcomS}) which parallels (\ref{WIDoff})
and (\ref{WIKoff}). \\
A superficial look at the WI's (\ref{WIDoff}) and (\ref{WIKoff}) says that
the Green functions are invariant under dilational and special conformal 
transformations once those are modified to include transformations of
the external field $\lambda$. In which sense can this be reconciled with
(\ref{dilcomS}) and (\ref{confcomS}) telling that the S-matrix of
$\phi^4$-theory is anomalous, i.e.\ the respective
charges do not commute with the $S$-operator? The statement on the charges
is obviously correct.
 We are, however allowed to rewrite (\ref{dilcomS}) and (\ref{confcomS})
as

\begin{align}
 \Bigg\{i[D,\,\, \bullet \, \, ] - \lim_{\lambda\rightarrow
 \textnormal{const}} \sum^\infty_{k=1}\limits \int dx\, \hat \beta^{(k)}_\lambda \lambda^{k+1}(x) \frac{\delta}{\delta \lambda(x)}\Bigg\} S &= 0, \label{conservcharge1}\\
 \Bigg\{i[K_\mu,\,\, \bullet \,\, ] - \lim_{\lambda\rightarrow\textnormal{const}}\sum^\infty_{k=1}\limits \int dx\; 2x_\mu\, \hat \beta^{(k)}_\lambda \lambda^{k+1}(x) \frac{\delta}{\delta \lambda(x)}\Bigg\} S &= 0, \label{conservcharge2}
\end{align}
where we may interpret the terms containing the $\beta$-function as
a contribution to the respective charge. The first term acts as it should 
as an operator in Hilbert space, the second term acts as it can on the 
operator $S$. This is perfectly legitimate
in a quantum field theory which depends on an external field; there it
is standard that an $S$-operator not only depends on quantum fields,
which propagate, but also on classical fields which do not propagate.
Hence also a charge operator may depend on classical fields. 
Without the external field $\lambda(x)$ the charges $D$ and $K_\mu$
could only change via one-particle singularities in the WI's. However
terms non-linear in the quantum field do not cause such singularities
in perturbation theory. And this is why the homogeneous WI version of 
the situation seduces to talk about symmetry which is, however only 
correct after noticing that the changes for $\lambda$ can indeed be
understood as being changes of the charges, which can be easily read
off from (\ref{conservcharge1}), (\ref{conservcharge2}). The operator
within curly brackets may be interpreted as a derivation, where the
two terms act on their respective
spaces, the whole space being $\textnormal{Hilbert} \bigotimes
\{\textnormal{external fields\}}$. 
 
Taking into account that both terms of the left-hand-sides of
(\ref{conservcharge1}), (\ref{conservcharge2}) are changes we may
write the complete
symmetry transformations for the $S$-operator to first order in the
transformation parameters as

\begin{align}
 \Bigg\{I+\epsilon\Bigg(
i[D,\,\, \bullet \, \, ] - \lim_{\lambda\rightarrow\textnormal{const}} \sum^\infty_{k=1}\limits \int dx\, \hat \beta^{(k)}_\lambda \lambda^{k+1}(x) \frac{\delta}{\delta \lambda(x)}\Bigg)\Bigg\} S &= S, \label{conservcharge3}\\
 \Bigg\{I+\alpha^\mu\Bigg(
i[K_\mu,\,\, \bullet \,\, ] - \lim_{\lambda\rightarrow\textnormal{const}} \sum^\infty_{k=1}\limits \int dx\; 2x_\mu\, \hat \beta^{(k)}_\lambda \lambda^{k+1}(x) \frac{\delta}{\delta \lambda(x)}\Bigg)\Bigg\} S &= S, \label{conservcharge4}
\end{align}
The charges $D$ and $K_\mu$ are not multiplicatively renormalized when going
from classical approximation (no loop, $\beta=0$) to non-trivial loop order,
but acquire an additive change via the $\beta$-function terms.

\section{$\beta$-function identifies changes of spacetime}
In this section we would like to show that the transformation laws for the
$S$-operator under dilatations and special conformal transformations
respectively
admit the definition of an $S$-operator on a spacetime which is obtained 
from standard Minkowski by performing the (infinitesimal) dilatation,
resp.\ special conformal transformation with parameters $\epsilon$ resp.\
$\alpha_\mu$.

For dilatations we have found the relation

\begin{align}
 i\epsilon [D,S(\lambda)] = \epsilon \sum^\infty_{k=1}\limits \int dx\, \hat \beta_\lambda^{(k)} \lambda^{k+1} \frac{\delta}{\delta \lambda(x)}\, S (\lambda)
 \label{eq:dila}
\end{align}

Let us recall how one finds the transformation law of a scalar field when
performing an infinitesimal translation $$x_\mu \rightarrow x_\mu - a_\mu$$ Requiring that the transformed field at the new coordinate be the same as the
old field at the old coordinate one finds $$\phi(x) \rightarrow \phi(x+a) = \phi(x) + a^\mu \pa_\mu \phi(x)$$
In the same sense we write the S-matrix in a transformed space, dilated
Minkowski, with an effective coupling
\begin{equation}
\lambda_\mathrm{eff} = \lambda + \epsilon\sum_k \hat\beta^{(k)} \lambda^{k+1},
\end{equation}
obtaining
\begin{align}
 S(\lambda) \rightarrow &  S (\lambda - \epsilon \sum_k \hat\beta^{(k)} \lambda^{k+1}) \\
 & = \sum_n S^{(n)} (\lambda - \epsilon \sum_k \hat\beta^{(k)} \lambda^{k+1})
\end{align}

Expanding this order by order in $\hbar$ we get

\begin{align}
 S^{(0)} & = S^{(0)}(\lambda) \\
 S^{(1)} & = S^{(1)}(\lambda) - \epsilon \int dx\, \hat\beta^{(1)} \lambda^2 \frac{\delta}{\delta \lambda (x)} S^{(0)}(\lambda)\\
 S^{(2)} & = S^{(2)}(\lambda) - \epsilon \int dx\, \hat\beta^{(1)} \lambda^2 \frac{\delta}{\delta \lambda (x)} S^{(1)}(\lambda) - \epsilon \int dx\, \hat\beta^{(2)} \lambda^3 \frac{\delta}{\delta \lambda (x)} S^{(0)}(\lambda)\\
 S^{(3)} & = S^{(3)}(\lambda) - \ldots \\
 & \vdots
\end{align}

Expanding the S-matrix in (\ref{eq:dila}) and omitting terms $\propto \epsilon^2$, i.e. $S(\lambda - \epsilon \sum_k \hat\beta^{(k)} \lambda^{k+1} ) \rightarrow S(\lambda)$, we get

\begin{align}
 i\epsilon[D,S(\lambda - \epsilon \sum_k \hat\beta^{(k)} \lambda^{k+1})] & = i\epsilon [D,\sum_n S^{(n)}(\lambda)] \\
 & = \epsilon \sum_k\limits \int dx\, \hat \beta^{(k)} \lambda^{k+1} \frac{\delta}{\delta \lambda(x)} \sum_n\limits S^{(n)} (\lambda)
\end{align}

Performing the transformation we have

\begin{align}
 S(\lambda - \epsilon \sum_k \hat\beta^{(k)} \lambda^{k+1}) +i\epsilon[D,S(\lambda - \epsilon \sum_k \hat\beta^{(k)} \lambda^{k+1})] = S(\lambda),
\end{align}
or by a shift 
\begin{align}
 S(\lambda) +i\epsilon[D,S(\lambda)] = S(\lambda_{\textnormal{eff}})
 \label{dilspaceS},
\end{align}

where the right-hand-side can be understood as an $S$-operator on a
flat space with metric $g_{\mu\nu} = (1-2\epsilon)\eta_{\mu\nu}$.
Here $\epsilon$ is the parameter determining the dilatation in question.
The non-vanishing $\beta$-function triggers the transition to this space.
If it were to vanish, dilatations would be realizable on the original
Minkowski space as a true symmetry. We could not detect the transformation.
Via the $S(\lambda_{\mathrm{eff}})$ and its non-trivial dependence on
$\epsilon$ we can however spot it. So, clearly one cannot identify the
underlying spacetime per se, out of nothing, but the {\sl change} from
a fourdimensional Minkowski space to a fourdimensional dilated space is
characterized by the $\beta$-function.

The case of special conformal transformations can be dealt with in
complete analogy,
\begin{equation}\label{confspaceS}
 S(\lambda) +i[\alpha^\mu K_\mu,S(\lambda)] =
                                       S(\lambda_{\textnormal{eff}}),
\end{equation}
where now $\lambda_{\textnormal{eff}}$ is given by
\begin{equation}
\lambda^{\textnormal{conf}}_{\textnormal{eff}} =
      \lambda + 2x^\mu\alpha_\mu \sum_k \hat\beta^{(k)} \lambda^{k+1}. 
\label{effcoupconf}
\end{equation}      
As above $\alpha$ is the infinitesimal parameter of the respective
special conformal transformation.\\
The metric of the transformed space reads to first order in $\alpha$
\begin{equation}
g_{\mu\nu} = (1-4x^\rho \alpha_\rho)\eta_{\mu\nu}.
\end{equation}
The $x$-dependence of effective coupling and metric have an interesting
effect: ``constant'' coupling is now to be understood as ``conformally
constant'', i.e.\ relative to the new metric!\\
Again the same comment applies as in the case of dilatations: due to
the non-vanishing $\beta$-function we can identify the (conformally)
transformed spacetime departing from ordinary Minkowski space.

\section{Discussion and conclusions}
In the present paper we treat two topics. The first one concerns
the change of the $S$-matrix under dilatations and special conformal
transformations of the massless $\phi^4$ theory in the context of perturbation
theory. For dilatations Zimmermann \cite{Zimmermannphi4} has obtained
the respective result (\ref{RGequat}) in an  axiomatic setting. We arrive
perturbatively at the same, (\ref{dilcomS}), by rendering the coupling
local and using
the fact that with the help of local $\lambda$ all dilatational anomalies
can be absorbed into a homogeneous WI (\ref{homDWI}) \cite{KrausSibold2}.
As long as $\lambda$ is local there is at every vertex a non-vanishing
external momentum,
hence one can go on-shell without meeting an infrared divergence.
Following Zimmermann we assume that the non-perturbative theory is linked
to the perturbative version in the sense of an asymptotic expansion.
We know therefore from \cite{Zimmermannphi4} that no infrared divergences
can arise in the limit of constant coupling, once we realize
perturbatively the same normalization conditions as employed by Zimmermann,
which we do.
The amputated on-shell Green functions are thus finite. Single diagrams
may be infrared divergent, but these divergences cancel in the sum which
represents the Green function. And, indeed, in the one-loop approximation
which is given by just one diagram, no infrared divergence shows up.\\
For special conformal transformations we find the analogous result,
(\ref{confcomS}). By the moment construction the special conformal
transformations are closely related to the dilatations, hence the
same cancellation of infrared divergences takes place, as the one-loop
approximation shows explicitly.\\
The learned reader might object that the $\phi^4$-theory is trivial.
However, rigorous proofs of triviality exist only for dimensions
strictly smaller or strictly larger than four. For exactly four
dimensions no such proof seems to be available.

As second topic we discuss an application of these results
for the $S$-matrix. We show that upon introduction of a suitable effective
coupling an $S$-matrix can be defined which signals the underlying
transformed spacetime which is obtained from Minkowski space by
dilatation, special conformal transformation respectively (\ref{dilspaceS}),
(\ref{confspaceS}). Crucial is here the fact that the $\beta$-function
does not vanish. In the case of dilatations this effective coupling
agrees with the standard running coupling obtained from the renormalization
group equation. The effective coupling for the special conformal case
(\ref{effcoupconf})
is particularly interesting because it depends explicitly on $x_\mu$.
Constant coupling then means conformally constant!\\
In our understanding we contribute with this result to a presently ongoing
general program: the construction of quantum field theories on non-trivial
spacetimes \cite{Dappiaggi1}, \cite{Dappiaggi2}, \cite{Pinamonti}.
Beginning perhaps with Wald \cite{Wald79} it gradually became clear how
to define $S$-matrices on spacetimes which are globally hyperbolic and
asymptotically flat. Here we present an explicit and non-trivial example:
the perturbatively non-trivial $S$-matrix of $\phi^4$-theory in Minkowski
space gets translated to a dilated, resp.\ conformally (flat) spacetime
via the effective coupling. If it is possible to go to finite
transformations one will obtain (in the conformal case) a space which is
curved, conformally and asymptotically flat. If one succeeds to find the 
corresponding expression for the $S$-matrix (i.e.\ one integrates the
effective coupling) one has a non-trivial $S$-matrix on such a space.\\
The axioms which hold for $S(\lambda)$ on Minkowski space get via
$S(\lambda^{\mathrm{conf}}_{\mathrm{eff}})$ translated on the transformed
space, wherefrom its explicit realization can be read off, notably that
of locality, i.e.\ causality.\\
$\beta$-functions arise from Zimmermann identities among normal products.
The Zimmermann coefficients encode the information on anomalies, hence
when interpreted as we propose also on spacetimes.\\
Which extensions of these results can one expect in other theories?
Of utmost interest are gauge theories. There, however one has to cope
with the fact that gauge fixing is not conformally invariant. Therefore
the identification of the physical subspace of the entire Fock space
is technically non-trivial. If the $S$-matrix exists, i.e.\ in models with 
complete breakdown of the gauge symmetry, one has to separate the soft
breaking of conformal invariance from hard breaking. But then one should
find essentially similar results as presented here. This expectation is
based on the fact, that the $\beta$-functions can be constructed as
gauge independent quantities. If the $S$-matrix does not exist (e.g.
\ in pure QCD) one has to consider other observables, the energy-momentum
tensor being a prime candidate. Here one has to check whether other
anomalies than those related to the $\beta$-function come into play.
As far the identification of an underlying spacetime is concerned
we expect in any case an analogous result to the above: anomalies of
geometric symmetries identify respective spacetimes.\\
What about anomalies
of internal symmetries? Here we expect a change of geometry of the
internal space. The relevant anomaly coefficients are, if properly
constructed,
also gauge independent, hence physical. The geometry of the internal
space might get a non-trivial physical meaning. These questions certainly
deserve further investigation.
 
{\bf Acknowlegements}\\
We are grateful to Burkhard Eden for helpful discussions.\\
This paper is dedicated to Wolfhart Zimmermann at the occasion of his
82nd birthday.

\textbf{\Large Appendix: one-loop approximation}

In this appendix we calculate explicitly the one-loop contribution
to the scattering amplitude of two particles going into two particles
and the conformal transformation of this process. There are two
diagrams contributing to this S-matrix element $S^{(1)}_{2,2}$:
they involve four fields and the outcome is equivalent to the
four-point vertex function $\Gamma^{(1)}_4$ at the same loop order. 

\begin{figure*}[h]
 \includegraphics{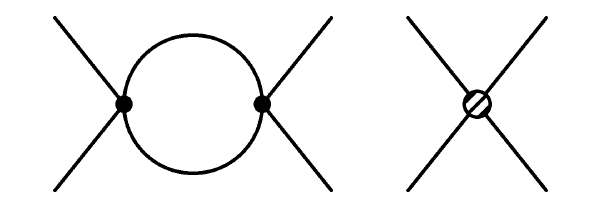}
\end{figure*}

We expect a contribution to the breaking of conformal invariance as
shown in \cite{KrausSibold2}, clearly for constant coupling. Since the
counterterm behaves trivially under special conformal transformations we
can omit its treatment. Then the relevant relation for $S^{(1)}_{2,2}$ is
\begin{align} 
 S^{(1)}_{2,2} & = \Sigma Z^{(1)}_{2,2}[J]_{|J=0,s=1} \\
 \Sigma & = \exp \{X\} = \exp \left\{ \int dx\, \phi_\mathrm{in}(x)\, r^{-1}\, \Box_x \frac{\delta}{\delta J(x)} \right\} 
\end{align}
\vspace{-0.8cm}
\begin{multline}
 Z^{(1)}_{2,2}[J] = \left( -i \frac{\lambda}{4!} \right)^2 \cdot 36 \cdot \int dz_1 \, dz_2 \, \left( \Delta_c(z_1 - z_2) \right)^2 \left( \int d\xi \, \Delta_c(z_1 - \xi) i J(\xi) \right)^2 \times \\ 
 \times \left( \int d\xi \, \Delta_c(z_2 - \xi) i J(\xi) \right)^2 \cdot Z_0
\end{multline}
By applying the Ward-Operator $W^K_\mu$ we first regain (\ref{chargeScom}). (Wick-dots are omitted)
\begin{align}
[\Sigma, W^K_\mu] & = [X,W^K_\mu] \Sigma \\
 & =  \int dx \, \phi_\mathrm{in}(x) \, r^{-1} \, \Box_x \, \delta^K_\mu (x,d=1)\frac{\delta}{\delta J(x)} \, \Sigma \\
 & = - \int dx \, \delta^K_\mu (x,d=1) \phi_\mathrm{in} (x) \, r^{-1} \, \Box_x \, \frac{\delta}{\delta J(x)} \, \Sigma \\
 & = - [K_\mu,\Sigma]
\end{align}
In the one-loop approximation the wave function renormalization constant $r=1$, hence the full expression of the transformed matrix element reads
\begin{multline}
 [W^K_\mu, \,S^{(1)}_{2,2}] = \frac{\lambda^2}{24} \int dz_1 \, dz_2 \, \Big( \phi_\mathrm{in}(z_1) \, \delta^K_\mu(z_1,d=1)\phi_\mathrm{in}(z_1) \, \phi_\mathrm{in}^2 (z_2) + \phi_\mathrm{in}^2 (z_1) \phi_\mathrm{in} (z_2) \\
 \delta^K_\mu(z_2,d=1) \phi_\mathrm{in}(z_2) \Big) \, \Delta_c^2
 (z_1 - z_2).
\end{multline}
In order to simplify the calculation we use the following relations
\begin{align}
 \phi_\mathrm{in}(x) \delta^K_\mu(x,d=1) \phi_\mathrm{in}(x) & = \frac{1}{2} \delta^K_\mu(x,d=2) \phi_\mathrm{in}^2 (x) \\
 \int \Delta_c^2(x) \delta^K_\mu(x,d=2) \phi_\mathrm{in}^2(x) & = -\int \phi_\mathrm{in}^2(x) \delta^K_\mu (x,d=2) \Delta_c^2(x)
\end{align}
and obtain
\begin{multline}
  [W^K_\mu, \,S^{(1)}_{2,2}] = -\frac{\lambda^2}{48} \int dz_1 \, dz_2\, \phi_\mathrm{in}^2(z_1) \phi_\mathrm{in}^2(z_2) \left(\delta^K_\mu(z_1 , d=2) + \delta^K_\mu(z_2,d=2)\right) \Delta_c^2(z_1 - z_2)
\end{multline}
Here
$\Delta^2_c$ is identical with $\Gamma^{(1)}_4$ and becomes well-defined
only after we have specified a renormalization scheme, because it is
logarithmically divergent by power counting. We choose the BPHZL-scheme
with the propagator as given in (\ref{propagator}) and one subtraction at $p=0$
and $s=0$, leading to
\begin{multline} 
 \Delta_c^2(z_1 - z_2) = \int \frac{dp}{(2\pi)^4} e^{-ip(z_1 - z_2)}  \int \frac{dk}{(2\pi)^4} \Bigg( \frac{i}{k^2 - M^2(s-1)^2 +
 i\varepsilon_\mathrm{Z}} \times \\
 \times \frac{i}{(p-k)^2 - M^2(s-1)^2 + i \varepsilon_\mathrm{Z}} 
  - \frac{i^2}{(k^2 - M^2 + i\varepsilon_\mathrm{Z})^2} \Bigg).
\end{multline}
With the Zimmermann-$\varepsilon_\mathrm{Z}$ this expression is
absolutely convergent.
We translate $\delta^K_{\mu}(\,.\,,d=2)$ via $\exp\{ip(z_1-z_2)\}$ into 
momentum space, where monomials of dimension $-1$ result, i.e.\ 
derivatives with respect to the external momentum $p$. Hence the
subtraction term vanishes and the first term becomes convergent in its
own right. \\
We introduce Feynman parameters, which help evaluating the integrals, and obtain
\begin{align}
 [W^K_\mu, \,S^{(1)}_{2,2}] = - \frac{i\lambda^2}{24} \frac{\beta_\lambda^{(1)}}{3} \int dz_1 \, 2z_{1,\mu} \, \phi_\mathrm{in}^4(z_1)
\end{align}
Until this stage of the calculation we did not consider symmetries of the
specific diagram with respect to the external lines. Noticing that $s-$,
$t-$, $u-$channel give the same contribution we have a factor $3$ and the
final result is
\begin{align}
 [W^K_\mu, \,S^{(1)}_{2,2}] = -\frac{i \lambda^2}{4!} \int dx\, 2x_\mu\, \beta_\lambda^{(1)} \,:\! \phi_\mathrm{in}^4(x):
\end{align}
Here we reintroduced the Wick dots. We note that we did not come across
any infrared divergence. This is result for the $S$-operator used in
the main text.

\providecommand{\href}[2]{#2}\begingroup\raggedright\endgroup

\end{document}